\documentclass[
 reprint, groupedaddress,
 amsmath,amssymb,
 nofootinbib,
 nobibnotes,
 aps,
 prl,
 a4paper]{revtex4-2}
\pdfoutput=1 %

\usepackage{braket}
\usepackage{amsmath,amssymb,amsfonts,mathrsfs}
\usepackage{amsthm}
\usepackage{mathtools}
\usepackage{graphicx}
\usepackage{xcolor}
\usepackage{framed}
\usepackage{silence}
\usepackage[colorlinks=true,
            allcolors=blue]{hyperref}

\usepackage[capitalize,nameinlink]{cleveref}
\usepackage{comment}
\usepackage{amsmath,tikz}
\usetikzlibrary{arrows.meta}

\theoremstyle{definition}

\theoremstyle{remark}

\newcommand{\tr}[1]{\operatorname{Tr}\!\left[#1\right]}

\usepackage{dsfont}
\usepackage{enumitem}

\renewcommand{\le}{\leqslant}

\renewcommand{\geq}{\geqslant}
\renewcommand{\leq}{\leqslant}

\begin{document}

\title{Witnessing Quantum Bayesian Inference beyond Classical Learning}

\author{Francesco Buscemi}
\email{buscemif@gmail.com}
\affiliation{Department of Mathematical Informatics, Nagoya University, Furo-cho Chikusa-ku, Nagoya 464-8601, Japan}

\begin{abstract}
Can an agent's predictions reveal whether its reasoning is based on a classical or a quantum model? Using known results on matrix factorizations, here we show that quantum Bayesian retrodiction can produce one-step-ahead forecasts incompatible with every classical Bayesian explanation based on learning about a fixed but unknown sampling law from conditionally independent observations. The separation has a sharp threshold in the number of outcomes: for any measurement with at most four outcomes, the quantum predictions admit a classical realization, irrespective of the prior state and the finite Hilbert-space dimension, whereas a five-outcome measurement on a single qubit already violates an explicit classical bound. This bound holds for arbitrary classical latent states, prior probabilities, and sampling laws, and involves only the agent's initial and updated predictions. Thus, in this setting, quantum Bayesian retrodiction is strictly more expressive than classical Bayesian learning, and this difference can be witnessed without observing or specifying the agent's internal description.
\end{abstract}
\maketitle

\noindent {\em Introduction.}---Drawing with replacement from an urn
of unknown composition provides a simple model of statistical
learning~\cite{feller1968,jaynes_2003}. An observation changes our
information about the urn, and hence our predictions for subsequent
draws, while leaving the urn's composition unchanged.
This is the classical learning framework recently considered by Fudenberg
and Lanzani~\cite{fudenberg2026forecasts}, who characterize when an
agent's initial and updated forecasts admit such a Bayesian
explanation. We take this framework as the classical benchmark
for the comparison developed here.

To make this precise, consider a family of idealized urns labeled by $z\in Z$, containing balls whose species are labeled by $x\in X$, where both sets are finite but otherwise arbitrary. Each urn has a known composition, represented by a probability distribution $u(x| z)$, while our initial uncertainty about which urn is used is represented by a prior distribution $r(z)$. The same urn is retained for successive draws, with the ball replaced and the contents mixed after each draw. Thus, conditional on the urn's identity, the draws are independent and identically distributed.

Suppose that the first draw gives a ball of species $x$. Bayes' theorem updates our uncertainty about the urn according to $r(z| x)\propto r(z)u(x| z)$. Since the composition of each candidate urn remains unchanged, our \textit{one-step-ahead forecast} is obtained by averaging the same sampling distributions under the updated prior, namely, $T(y| x)=\sum_z r(z| x)u(y| z)$. Consequently, the joint probability distribution of the two draws is
\begin{equation}
T_{xy}=\sum_z r(z)u(x| z)u(y| z).
\label{eq:classical-joint}
\end{equation}
In other words, correlations between successive draws arise entirely from uncertainty about a fixed sampling law. No restriction is imposed on the number of candidate urns or on their compositions.

Quantum generalizations of Bayes' rule have recently been investigated from several perspectives~\cite{leifer-spekkens-2013-causally-neutral,Parzygnat2023axiomsretrodiction,parzygnat-fullwood-2023-bayes,bai-2025-q-bayes-minimum-change,karmakar2026bayesian} (see also the references therein). In particular, the \textit{quantum minimum-change principle} provides a variational prescription for retrodiction by minimizing the change in the description of the input-output process subject to the observed evidence~\cite{bai-2025-q-bayes-minimum-change,Kuang_2026}. For a prior quantum state $\gamma$ and a POVM $\{P_x\}_{x\in X}$, observing an outcome $x$ with probability $p_x=\operatorname{Tr}(\gamma P_x)>0$ leads to the retrodictive assignment $\gamma_x\propto\sqrt{\gamma}P_x\sqrt{\gamma}$. Using this state to predict the same POVM gives $T(y| x)=\tr{P_y\gamma_x}$, and hence the probability table
\begin{equation}
T_{xy}=p_xT(y| x)
=\tr{\sqrt{\gamma}P_x\sqrt{\gamma}P_y}.
\label{eq:quantum-joint}
\end{equation}
When $\gamma$ and all the effects $P_x$ are mutually commuting, simultaneous diagonalization recovers Eq.~\eqref{eq:classical-joint}. The urn model is therefore the commutative instance of the same inferential construction.

The question then arises whether the forecasts obtained from quantum
Bayesian retrodiction always admit an explanation within this
classical learning framework. More precisely, does every table of
the form~\eqref{eq:quantum-joint} admit a classical
representation~\eqref{eq:classical-joint}, possibly involving an
arbitrarily large latent space $Z$?

In this work, we show how known results on matrix factorizations
imply a separation between quantum Bayesian retrodiction and
classical Bayesian learning. More precisely, minimum-change quantum
retrodiction cannot, in general, be represented as Bayesian learning
about a fixed sampling law from conditionally independent observations.
The same matrix results imply a sharp threshold in the number of
outcomes: every quantum retrodictive table with at most four outcomes
admits a classical realization, independently of the finite
Hilbert-space dimension, whereas a five-outcome measurement on a
qubit already produces predictions that no classical model of the
form~\eqref{eq:classical-joint} can reproduce.

\  

\noindent {\em Classical and quantum realizations.}---The data to be explained are the agent's initial probabilities $p_x$ and its one-step-ahead forecasts $T(y|x)$ after observing $x$. Together, they determine the table $T_{xy}=p_xT(y|x)$. No further draw need actually be performed: in the quantum construction, the updated forecast is obtained from the retrodictive state $\gamma_x$ through the Born rule. We ask whether these forecasts admit a classical explanation, irrespective of the model from which they were obtained. Throughout, such an explanation means a \textit{conditionally i.i.d.\ representation} of the form~\eqref{eq:classical-joint}, with the same conditional sampling law $u(\cdot|z)$ for both draws\footnote{If different conditional sampling laws $u_1$ and $u_2$ are allowed for the two draws, classical realizability becomes a moot point, since every joint distribution admits a representation $T_{xy}=\sum_z r(z)u_1(x|z)u_2(y|z)$. Indeed, one may take $z=(a,b)$, $r(a,b)=T_{ab}$, $u_1(x|a,b)=\delta_{x,a}$, and $u_2(y|a,b)=\delta_{y,b}$.}.

To characterize classical realizability, notice first that every classical model gives a factorization $T=BB^{\mathsf T}$ with $B$ entrywise nonnegative: setting $B_{xz}=\sqrt{r(z)}u(x|z)$ in Eq.~\eqref{eq:classical-joint} gives precisely this form, where superscript $\mathsf T$ denotes transposition. Conversely, suppose that a normalized table $T$ admits such a factorization. Discarding the zero columns of $B$, define $s_z=\sum_x B_{xz}$, $r(z)=s_z^2$, and $u(x|z)=B_{xz}/s_z$. The distributions $u(\cdot|z)$ are normalized by construction, while $\sum_z r(z)=\sum_{xy}T_{xy}=1$; substituting them into Eq.~\eqref{eq:classical-joint} therefore recovers $T$. Thus, a classical realization exists if and only if $T$ is a {\em completely positive matrix}, namely, a matrix admitting a factorization $T=BB^{\mathsf T}$ with $B$ entrywise nonnegative. This is also the criterion underlying the characterization of Bayesian forecasts in Ref.~\cite{fudenberg2026forecasts}. Here, complete positivity is a property of a numerical matrix, distinct from complete positivity of a quantum map.

The quantum case admits a parallel characterization, with nonnegative vectors replaced by positive semidefinite operators. Indeed, setting $Q_x=\gamma^{1/4}P_x\gamma^{1/4}$ in Eq.~\eqref{eq:quantum-joint} gives
\begin{equation}
T_{xy}=\tr{Q_xQ_y},\qquad Q_x\succeq0,
\label{eq:positive-gram}
\end{equation}
where $\succeq$ denotes the Loewner order. Conversely, every normalized table of the form~\eqref{eq:positive-gram} can be generated by a suitable quantum prior and POVM. To see this, start with the positive semidefinite operators $Q_x$ appearing in that representation and set $S=\sum_xQ_x$. Normalization of the table gives $\tr{S^2}=\sum_{xy}\tr{Q_xQ_y}=1$. We can therefore choose $\gamma=S^2$ as the prior state and, on the support of $S$, define $P_x=S^{-1/2}Q_xS^{-1/2}$. These operators form a POVM, since each $P_x$ is positive semidefinite and $\sum_xP_x=I$ on that support. Moreover, $\gamma^{1/4}P_x\gamma^{1/4}=Q_x$, so this prior and POVM reproduce the original table through Eq.~\eqref{eq:quantum-joint}.

Matrices admitting positive semidefinite Gram representations of the form~\eqref{eq:positive-gram} are called {\em completely positive semidefinite} matrices~\cite{laurent-piovesan-2015,fawzi-2015-psd-rank}. The comparison between classical and quantum realizations thus leads to two matrix cones that also arise in the study of classical and quantum correlations~\cite{prakash-2018-cpsd-rank}. To apply the results on these cones, however, one point requires clarification: the standard definition uses real symmetric factors, whereas the operators $Q_x$ in our construction may have complex entries. As already noted in Ref.~\cite{fawzi-2015-psd-rank}, this difference does not change the class of admissible tables when the finite dimension is unrestricted.

Indeed, write $Q_x=A_x+iB_x$, with $A_x$ and $B_x$ real. Hermiticity gives $A_x^{\mathsf T}=A_x$ and $B_x^{\mathsf T}=-B_x$, so
\begin{equation}
\mathcal R(Q_x)=\frac{1}{\sqrt2}
\begin{pmatrix}
A_x & -B_x\\
B_x & A_x
\end{pmatrix}
\label{eq:realification}
\end{equation}
is real symmetric and positive semidefinite: its quadratic form on a real vector $(a,b)$ equals $(a+ib)^\dagger Q_x(a+ib)/\sqrt2\geq0$. Moreover, $\tr{\mathcal R(Q_x)\mathcal R(Q_y)}=\tr{Q_xQ_y}$. Hence, every complex representation can be replaced by a real one in twice the dimension, without changing $T$.

Classical and quantum retrodictive tables are therefore exactly the normalized completely positive and completely positive semidefinite matrices, respectively, with finite-dimensional realizations understood throughout. Notice that every classical factorization is a particular quantum factorization, obtained by taking $Q_x=\operatorname{diag}(B_{xz})_z$. The question is whether the converse holds: when does a table admitting a Gram representation by positive semidefinite operators also admit one by nonnegative vectors?

\  

\noindent {\em Classical realizability with at most four outcomes.}---Let us first make explicit which properties of a retrodictive table are relevant to the comparison. Writing $n=|X|$, the table $T$ is a real $n\times n$ matrix with entries $T_{xy}=p_xT(y|x)\geq0$ and $\sum_{xy}T_{xy}=1$. In both the classical and quantum constructions, it is symmetric, namely, $T_{xy}=T_{yx}$, and its row and column marginals coincide with the initial outcome probabilities: $\sum_yT_{xy}=\sum_yT_{yx}=p_x$.

There is, moreover, a constraint beyond nonnegativity and symmetry. By Eq.~\eqref{eq:positive-gram}, every quantum retrodictive table is positive semidefinite, since
\begin{equation}
\sum_{xy}v_x^*T_{xy}v_y
=\left\|\sum_xv_xQ_x\right\|^2_2\geq0\;,
\label{eq:table-positivity}
\end{equation}
where superscript $*$ denotes complex conjugation and $\|\cdot\|_2$ the Hilbert--Schmidt norm, for every complex vector $v$. The same condition is necessary for a classical realization, as follows from $T=BB^{\mathsf T}$. The question is whether these readily verified properties already suffice to ensure the existence of a classical explanation.

For $n\leq4$, the answer is affirmative. A theorem of Maxfield and Minc states that every real symmetric positive semidefinite matrix of order at most four with nonnegative entries admits a factorization $T=BB^{\mathsf T}$ with $B$ entrywise nonnegative~\cite{maxfield-minc-1962}. Normalization then converts this factorization into a prior and a family of urn compositions, as shown above. Hence, every normalized table satisfying these conditions is classically realizable, as also stated for Bayesian forecasts in Corollary~2 of Ref.~\cite{fudenberg2026forecasts}. In particular, every quantum retrodictive table with $n\le 4$ outcomes admits a classical realization. In fact, at most $n$ candidate urns suffice. This conclusion holds in arbitrary finite Hilbert-space dimension.

The content of this statement is somewhat surprising when applied to an informationally complete four-outcome qubit measurement. Indeed, although the effects of such a measurement necessarily fail to commute, the corresponding retrodictive table admits an explanation in terms of classical Bayesian learning for every prior state. Thus, the distinction between quantum retrodiction and classical learning cannot be reduced to the distinction between commuting and noncommuting models.

For a concrete example, take the tetrahedral POVM $P_x=(I+\boldsymbol n_x\cdot\boldsymbol\sigma)/4$, where $\boldsymbol\sigma$ is the vector of Pauli matrices and the four unit vectors $\boldsymbol n_x$ point towards the vertices of a regular tetrahedron, so that $\boldsymbol n_x\cdot\boldsymbol n_y=-1/3$ for $x\ne y$. With the maximally mixed prior $\gamma=I/2$, the resulting table is
\begin{equation}
T_{xy}=\frac{1+\boldsymbol n_x\cdot\boldsymbol n_y}{16}
=\frac{1+2\delta_{xy}}{24}.
\label{eq:tetrahedral-table}
\end{equation}
An explicit classical realization uses four equally likely urns, $r(z)=1/4$, with compositions $u(x|z)=(1-1/\sqrt3)/4+\delta_{xz}/\sqrt3$. Each urn favors one species and assigns equal probabilities to the other three; substituting these distributions into Eq.~\eqref{eq:classical-joint} gives exactly Eq.~\eqref{eq:tetrahedral-table}. Thus, even informational completeness does not make the quantum origin of these forecasts distinguishable from a classical explanation.

\  

\begin{figure}[t]
\centering
\begin{tikzpicture}[
  font=\small,
  outcome/.style={circle,draw,fill=white,minimum size=6.5mm,
                  inner sep=0pt,line width=.7pt}
]
\def\pentR{1.9cm}
\foreach \j in {0,...,4}
  \coordinate (v\j) at ({90-72*\j}:\pentR);
\draw[black!65,line width=.8pt]
  (v0)--(v1)--(v2)--(v3)--(v4)--cycle;
\foreach \j in {0,...,4}
  \node[outcome] at (v\j) {$\j$};
\node[above=5mm]         at (v0) {$u_0$};
\node[right=5mm]         at (v1) {$u_1$};
\node[below right=3.5mm] at (v2) {$u_2$};
\node[below left=3.5mm]  at (v3) {$u_3$};
\node[left=5mm]          at (v4) {$u_4$};
\end{tikzpicture}
\caption{Five species arranged in a cycle, with drawing probabilities
$u_j$. An edge connects neighboring species; hence, the probability of
neighboring outcomes in two draws with replacement is
$f(u)=2\sum_{j=0}^4u_ju_{j+1}$, with indices understood modulo five.}
\label{fig:classical-bound}
\end{figure}
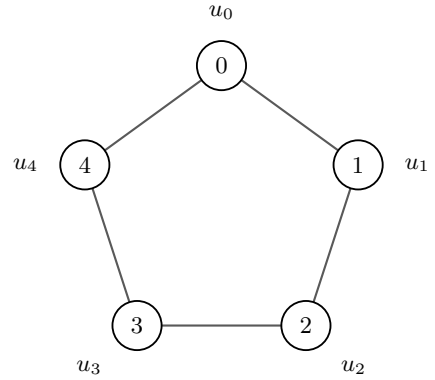

\noindent {\em A classical bound.}---To determine whether five outcomes suffice for a separation, we first look for a constraint that every classical learning model with five outcomes must obey. Label the outcomes by $j=0,\ldots,4$ and arrange the labels around a pentagon as in Fig.~\ref{fig:classical-bound}, with indices understood modulo five. We ask how likely two draws are to give neighboring labels, counting both possible orders. The pentagon specifies which pairs enter this probability; it imposes no restriction on the compositions of the candidate urns.

Let us begin with a single urn containing five possible species of balls, and denote by $u_j$ the probability of drawing species $j$. Since the draws are made with replacement, the probability of obtaining neighboring species, in either order, is $f(u)=2\sum_{j=0}^4u_ju_{j+1}$, with indices understood modulo five.

The bound $f(u)\leq1/2$ is the five-cycle instance of the Motzkin--Straus theorem~\cite{motzkin-straus-1965}, but in this case it follows from a simple inequality. Indeed, by cyclically relabeling the species, we may assume that $u_0$ is the largest probability. Replacing the term $u_3u_4$ in the sum by $u_3u_0$, and adding the nonnegative term $u_2u_4$, can only increase the sum. Hence,
\begin{align}
f(u)&=2(u_0u_1+u_1u_2+u_2u_3+u_3u_4+u_4u_0)\nonumber\\
&\leq2(u_0+u_2)(u_1+u_3+u_4)\leq\frac12.
\label{eq:single-urn-bound}
\end{align}
The last inequality follows because the two nonnegative factors in parentheses sum to one, so their product cannot exceed $1/4$. Equality is attained by an equal mixture of two neighboring species, for example, $u_0=u_1=1/2$ with all other probabilities zero.

Averaging over any prior preserves the bound. Hence every classical retrodictive table satisfies
\begin{equation}
\mathcal B_5(T):=2\sum_{j=0}^{4}T_{j,j+1}\leq\frac12.
\label{eq:classical-bound}
\end{equation}
The bound is attained even with uniform initial outcome probabilities: one can choose, with equal prior probability, among five urns, each containing an equal mixture of the two species at the ends of one edge. A violation of Eq.~\eqref{eq:classical-bound} therefore witnesses the impossibility of a classical realization of the form~\eqref{eq:classical-joint}.

\ 

\noindent {\em A minimal counterexample.}---We now show that quantum retrodiction can violate Eq.~\eqref{eq:classical-bound} using a single qubit. Take the prior $\gamma=I/2$ and the five-outcome POVM
\begin{equation}
P_j=\frac{I+\boldsymbol n_j\cdot\boldsymbol\sigma}{5},\qquad
\boldsymbol n_j=\left(\cos\frac{2\pi j}{5},0,
\sin\frac{2\pi j}{5}\right).
\label{eq:pentagon-povm}
\end{equation}
The five Bloch vectors lie on a common great circle, chosen here in the $xz$-plane. Hence, only $\sigma_x$ and $\sigma_z$ occur, and all effects are real symmetric matrices in the same basis. Each effect has eigenvalues $0$ and $2/5$, while $\sum_j\boldsymbol n_j=0$ ensures that $\sum_jP_j=I$. All outcomes have probability $p_j=1/5$, and observing $j$ gives the pure retrodictive state $\gamma_j=(I+\boldsymbol n_j\cdot\boldsymbol\sigma)/2$. Using this state to predict the same POVM gives
\begin{equation}
T_{jk}=\frac{1+\cos[2\pi(j-k)/5]}{25}.
\label{eq:pentagon-table}
\end{equation}
The matrix and its separating inequality are the normalized version of the example discussed in Sec.~8 of Ref.~\cite{fawzi-2015-psd-rank}; see also Sec.~3.2 of Ref.~\cite{laurent-piovesan-2015}, where the inequality is expressed through the Horn matrix. Here the table arises directly from quantum Bayesian retrodiction, and its failure to admit a classical realization is witnessed by
\begin{equation}
\mathcal B_5(T)=\frac{2}{5}\left(1+\cos\frac{2\pi}{5}\right)
=\frac{3+\sqrt5}{10}> 0.523>\frac12.
\label{eq:quantum-violation}
\end{equation}
Thus, no choice of prior, urn compositions, or number of candidate urns can reproduce these predictions.

\ 

\noindent {\em A universal upper bound.}---The pentagon witness also admits an upper bound valid for every quantum retrodictive table, independently of the Hilbert-space dimension. In fact, positive semidefiniteness and entrywise nonnegativity alone suffice. These properties define the larger cone of {\em doubly nonnegative} matrices, for which the optimal pentagon bound is known~\cite{burer-2009-dnn-cp}. Here we give a direct elementary proof.

With indices understood modulo five, write $F=2\sum_{j=0}^4T_{j,j+2}\geq0$ for the total probability assigned to nonneighboring distinct labels. Normalization then gives $\sum_jT_{jj}=1-\mathcal B_5(T)-F$. Choosing the vector with components $v_j=e^{4\pi i j/5}$, positive semidefiniteness and symmetry of $T$ imply
\begin{align}
0\leq v^\dagger Tv
&=\sum_jT_{jj}
+\cos\frac{4\pi}{5}\,\mathcal B_5(T)
+\cos\frac{8\pi}{5}\,F \nonumber\\
&=1-\frac{5+\sqrt5}{4}\,\mathcal B_5(T)
-\frac{5-\sqrt5}{4}\,F.
\end{align}
Since $F\geq0$, we obtain
\begin{equation}
\mathcal B_5(T)\leq\frac{4}{5+\sqrt5}
=1-\frac{1}{\sqrt5}<0.553\;.
\label{eq:universal-bound}
\end{equation}
Notice that this argument applies to the entire doubly nonnegative cone. It therefore provides only an upper bound for quantum retrodiction.

\ 

\noindent {\em Discussion.}---We have shown that an agent's predictions can, in certain situations, rule out every classical Bayesian explanation based on learning about a fixed sampling law. No access to the agent's internal description is required, and neither the latent variable $z$, nor the prior over it, nor the candidate sampling laws need be specified. Only the initial and updated predictions enter the comparison. In this precise sense, quantum Bayesian retrodiction is strictly more expressive than classical Bayesian learning within the specified model. Quantum realizability is established by the explicit construction, while the witness excludes classical explanations; the distinction concerns the inferential model, rather than the agent's physical implementation.

The use of retrodiction is central to this inferential interpretation. A measurement \textit{instrument} specifies the physical state change associated with an outcome, whereas Bayesian conditioning in the present learning model updates our information about the fixed latent state underlying the observations. These operations need not coincide. Indeed, allowing arbitrary measurement-induced state changes would also enlarge the classical comparison: a classical register can reproduce any joint table by starting with distribution $p_x$, reporting its value $x$, and then changing its value according to $T(y|x)$ before the next reading.

For a fair comparison with the Bayesian learning framework considered by Fudenberg and Lanzani~\cite{fudenberg2026forecasts}, we therefore adhered to the minimum-change principle, which is behind both classical and quantum updating, and singles out the retrodictive assignment $\gamma_x=\sqrt{\gamma}P_x\sqrt{\gamma}/p_x$ from the prior and the observed evidence~\cite{bai-2025-q-bayes-minimum-change,Kuang_2026}. The comparison thus concerns the forecasts obtained from this inferential prescription and their compatibility with learning about a fixed classical sampling law. It does not concern arbitrary changes of predictions caused by the physical action of a measuring device.

Within this inferential setting, several questions remain open. What consequences does this greater expressive power have for models of learning, and can it provide an advantage in concrete inference tasks? What is the maximal quantum value of $\mathcal B_5$, and which priors and measurements attain it? More generally, how do quantum violations of bounds obeyed by conditionally i.i.d.\ classical models depend on the Hilbert-space dimension, and how robust are they to noise? Can inequalities such as Eq.~\eqref{eq:classical-bound} be developed into operational witnesses of quantum reasoning? More broadly, the present approach suggests a programme for benchmarking and certifying quantum reasoning, in which the inferential capabilities of an agent are tested through its predictions alone. The aim is to make reasoning itself subject to certification, even when the agent's internal description remains inaccessible.

\

\noindent \textit{Acknowledgments.}---The author acknowledges support from MEXT Quantum Leap Flagship Program (MEXT QLEAP) Grant No.~JPMXS0120319794 and from JSPS KAKENHI Grants No.~23K03230 and~26K00621. ChatGPT (OpenAI, Astra 6 and Sol 6.1) was used, under the author's direction, to assist with literature searches, the examination of mathematical arguments, and the preparation and revision of the manuscript. The author checked all scientific statements, derivations, and references, and takes full responsibility for the content of this work.

\bibliography{library}

\begin{thebibliography}{15}%
\makeatletter
\providecommand \@ifxundefined [1]{%
 \@ifx{#1\undefined}
}%
\providecommand \@ifnum [1]{%
 \ifnum #1\expandafter \@firstoftwo
 \else \expandafter \@secondoftwo
 \fi
}%
\providecommand \@ifx [1]{%
 \ifx #1\expandafter \@firstoftwo
 \else \expandafter \@secondoftwo
 \fi
}%
\providecommand \natexlab [1]{#1}%
\providecommand \enquote  [1]{``#1''}%
\providecommand \bibnamefont  [1]{#1}%
\providecommand \bibfnamefont [1]{#1}%
\providecommand \citenamefont [1]{#1}%
\providecommand \href@noop [0]{\@secondoftwo}%
\providecommand \href [0]{\begingroup \@sanitize@url \@href}%
\providecommand \@href[1]{\@@startlink{#1}\@@href}%
\providecommand \@@href[1]{\endgroup#1\@@endlink}%
\providecommand \@sanitize@url [0]{\catcode `\\12\catcode `\$12\catcode `\&12\catcode `\#12\catcode `\^12\catcode `\_12\catcode `\%12\relax}%
\providecommand \@@startlink[1]{}%
\providecommand \@@endlink[0]{}%
\providecommand \url  [0]{\begingroup\@sanitize@url \@url }%
\providecommand \@url [1]{\endgroup\@href {#1}{\urlprefix }}%
\providecommand \urlprefix  [0]{URL }%
\providecommand \Eprint [0]{\href }%
\providecommand \doibase [0]{https://doi.org/}%
\providecommand \selectlanguage [0]{\@gobble}%
\providecommand \bibinfo  [0]{\@secondoftwo}%
\providecommand \bibfield  [0]{\@secondoftwo}%
\providecommand \translation [1]{[#1]}%
\providecommand \BibitemOpen [0]{}%
\providecommand \bibitemStop [0]{}%
\providecommand \bibitemNoStop [0]{.\EOS\space}%
\providecommand \EOS [0]{\spacefactor3000\relax}%
\providecommand \BibitemShut  [1]{\csname bibitem#1\endcsname}%
\let\auto@bib@innerbib\@empty
\bibitem [{\citenamefont {Feller}(1968)}]{feller1968}%
  \BibitemOpen
  \bibfield  {author} {\bibinfo {author} {\bibfnamefont {W.}~\bibnamefont {Feller}},\ }\href@noop {} {\emph {\bibinfo {title} {An Introduction to Probability Theory and Its Applications}}},\ \bibinfo {edition} {3rd}\ ed.,\ Vol.~\bibinfo {volume} {1}\ (\bibinfo  {publisher} {Wiley},\ \bibinfo {address} {New York},\ \bibinfo {year} {1968})\BibitemShut {NoStop}%
\bibitem [{\citenamefont {Jaynes}(2003)}]{jaynes_2003}%
  \BibitemOpen
  \bibfield  {author} {\bibinfo {author} {\bibfnamefont {E.~T.}\ \bibnamefont {Jaynes}},\ }\href {https://doi.org/10.1017/CBO9780511790423} {\emph {\bibinfo {title} {Probability Theory: The Logic of Science}}}\ (\bibinfo  {publisher} {Cambridge University Press},\ \bibinfo {year} {2003})\BibitemShut {NoStop}%
\bibitem [{\citenamefont {Fudenberg}\ and\ \citenamefont {Lanzani}(2026)}]{fudenberg2026forecasts}%
  \BibitemOpen
  \bibfield  {author} {\bibinfo {author} {\bibfnamefont {D.}~\bibnamefont {Fudenberg}}\ and\ \bibinfo {author} {\bibfnamefont {G.}~\bibnamefont {Lanzani}},\ }\bibfield  {title} {\bibinfo {title} {Forecasts and conditionally {I.I.D.} models}} (\bibinfo {year} {2026}),\ \bibinfo {note} {working paper, MIT Department of Economics (version of February 28, 2026), \url{https://economics.mit.edu/people/faculty/drew-fudenberg/working-papers}}\BibitemShut {NoStop}%
\bibitem [{\citenamefont {Leifer}\ and\ \citenamefont {Spekkens}(2013)}]{leifer-spekkens-2013-causally-neutral}%
  \BibitemOpen
  \bibfield  {author} {\bibinfo {author} {\bibfnamefont {M.~S.}\ \bibnamefont {Leifer}}\ and\ \bibinfo {author} {\bibfnamefont {R.~W.}\ \bibnamefont {Spekkens}},\ }\bibfield  {title} {\bibinfo {title} {Towards a formulation of quantum theory as a causally neutral theory of bayesian inference},\ }\href {https://doi.org/10.1103/PhysRevA.88.052130} {\bibfield  {journal} {\bibinfo  {journal} {Phys. Rev. A}\ }\textbf {\bibinfo {volume} {88}},\ \bibinfo {pages} {052130} (\bibinfo {year} {2013})}\BibitemShut {NoStop}%
\bibitem [{\citenamefont {Parzygnat}\ and\ \citenamefont {Buscemi}(2023)}]{Parzygnat2023axiomsretrodiction}%
  \BibitemOpen
  \bibfield  {author} {\bibinfo {author} {\bibfnamefont {A.~J.}\ \bibnamefont {Parzygnat}}\ and\ \bibinfo {author} {\bibfnamefont {F.}~\bibnamefont {Buscemi}},\ }\bibfield  {title} {\bibinfo {title} {Axioms for retrodiction: achieving time-reversal symmetry with a prior},\ }\href {https://doi.org/10.22331/q-2023-05-23-1013} {\bibfield  {journal} {\bibinfo  {journal} {{Quantum}}\ }\textbf {\bibinfo {volume} {7}},\ \bibinfo {pages} {1013} (\bibinfo {year} {2023})}\BibitemShut {NoStop}%
\bibitem [{\citenamefont {Parzygnat}\ and\ \citenamefont {Fullwood}(2023)}]{parzygnat-fullwood-2023-bayes}%
  \BibitemOpen
  \bibfield  {author} {\bibinfo {author} {\bibfnamefont {A.~J.}\ \bibnamefont {Parzygnat}}\ and\ \bibinfo {author} {\bibfnamefont {J.}~\bibnamefont {Fullwood}},\ }\bibfield  {title} {\bibinfo {title} {From time-reversal symmetry to quantum bayes' rules},\ }\href {https://doi.org/10.1103/PRXQuantum.4.020334} {\bibfield  {journal} {\bibinfo  {journal} {PRX Quantum}\ }\textbf {\bibinfo {volume} {4}},\ \bibinfo {pages} {020334} (\bibinfo {year} {2023})}\BibitemShut {NoStop}%
\bibitem [{\citenamefont {Bai}\ \emph {et~al.}(2025)\citenamefont {Bai}, \citenamefont {Buscemi},\ and\ \citenamefont {Scarani}}]{bai-2025-q-bayes-minimum-change}%
  \BibitemOpen
  \bibfield  {author} {\bibinfo {author} {\bibfnamefont {G.}~\bibnamefont {Bai}}, \bibinfo {author} {\bibfnamefont {F.}~\bibnamefont {Buscemi}},\ and\ \bibinfo {author} {\bibfnamefont {V.}~\bibnamefont {Scarani}},\ }\bibfield  {title} {\bibinfo {title} {Quantum {B}ayes' rule and {P}etz transpose map from the minimum change principle},\ }\href {https://doi.org/10.1103/5n4p-bxhm} {\bibfield  {journal} {\bibinfo  {journal} {Phys. Rev. Lett.}\ }\textbf {\bibinfo {volume} {135}},\ \bibinfo {pages} {090203} (\bibinfo {year} {2025})}\BibitemShut {NoStop}%
\bibitem [{\citenamefont {Karmakar}\ and\ \citenamefont {Parzygnat}(2026)}]{karmakar2026bayesian}%
  \BibitemOpen
  \bibfield  {author} {\bibinfo {author} {\bibfnamefont {P.}~\bibnamefont {Karmakar}}\ and\ \bibinfo {author} {\bibfnamefont {A.~J.}\ \bibnamefont {Parzygnat}},\ }\href@noop {} {\bibinfo {title} {Bayesian inference and retrodiction for faithful states on von {N}eumann algebras}} (\bibinfo {year} {2026}),\ \Eprint {https://arxiv.org/abs/2608.20001} {arXiv:2608.20001} \BibitemShut {NoStop}%
\bibitem [{\citenamefont {Kuang}\ \emph {et~al.}(2026)\citenamefont {Kuang}, \citenamefont {Torii},\ and\ \citenamefont {Buscemi}}]{Kuang_2026}%
  \BibitemOpen
  \bibfield  {author} {\bibinfo {author} {\bibfnamefont {J.}~\bibnamefont {Kuang}}, \bibinfo {author} {\bibfnamefont {K.}~\bibnamefont {Torii}},\ and\ \bibinfo {author} {\bibfnamefont {F.}~\bibnamefont {Buscemi}},\ }\bibfield  {title} {\bibinfo {title} {Quantum measurement retrodiction and entropic uncertainty relations},\ }\href {https://doi.org/10.1088/2058-9565/ae5fcd} {\bibfield  {journal} {\bibinfo  {journal} {Quantum Science and Technology}\ }\textbf {\bibinfo {volume} {11}},\ \bibinfo {pages} {025049} (\bibinfo {year} {2026})}\BibitemShut {NoStop}%
\bibitem [{\citenamefont {Laurent}\ and\ \citenamefont {Piovesan}(2015)}]{laurent-piovesan-2015}%
  \BibitemOpen
  \bibfield  {author} {\bibinfo {author} {\bibfnamefont {M.}~\bibnamefont {Laurent}}\ and\ \bibinfo {author} {\bibfnamefont {T.}~\bibnamefont {Piovesan}},\ }\bibfield  {title} {\bibinfo {title} {Conic approach to quantum graph parameters using linear optimization over the completely positive semidefinite cone},\ }\href {https://doi.org/10.1137/14097865X} {\bibfield  {journal} {\bibinfo  {journal} {SIAM Journal on Optimization}\ }\textbf {\bibinfo {volume} {25}},\ \bibinfo {pages} {2461} (\bibinfo {year} {2015})},\ \Eprint {https://arxiv.org/abs/1312.6643} {arXiv:1312.6643} \BibitemShut {NoStop}%
\bibitem [{\citenamefont {Fawzi}\ \emph {et~al.}(2015)\citenamefont {Fawzi}, \citenamefont {Gouveia}, \citenamefont {Parrilo}, \citenamefont {Robinson},\ and\ \citenamefont {Thomas}}]{fawzi-2015-psd-rank}%
  \BibitemOpen
  \bibfield  {author} {\bibinfo {author} {\bibfnamefont {H.}~\bibnamefont {Fawzi}}, \bibinfo {author} {\bibfnamefont {J.~a.}\ \bibnamefont {Gouveia}}, \bibinfo {author} {\bibfnamefont {P.~A.}\ \bibnamefont {Parrilo}}, \bibinfo {author} {\bibfnamefont {R.~Z.}\ \bibnamefont {Robinson}},\ and\ \bibinfo {author} {\bibfnamefont {R.~R.}\ \bibnamefont {Thomas}},\ }\bibfield  {title} {\bibinfo {title} {Positive semidefinite rank},\ }\href {https://doi.org/10.1007/s10107-015-0922-1} {\bibfield  {journal} {\bibinfo  {journal} {Mathematical Programming}\ }\textbf {\bibinfo {volume} {153}},\ \bibinfo {pages} {133} (\bibinfo {year} {2015})},\ \Eprint {https://arxiv.org/abs/1407.4095} {arXiv:1407.4095 [math.OC]} \BibitemShut {NoStop}%
\bibitem [{\citenamefont {Prakash}\ \emph {et~al.}(2018)\citenamefont {Prakash}, \citenamefont {Sikora}, \citenamefont {Varvitsiotis},\ and\ \citenamefont {Wei}}]{prakash-2018-cpsd-rank}%
  \BibitemOpen
  \bibfield  {author} {\bibinfo {author} {\bibfnamefont {A.}~\bibnamefont {Prakash}}, \bibinfo {author} {\bibfnamefont {J.}~\bibnamefont {Sikora}}, \bibinfo {author} {\bibfnamefont {A.}~\bibnamefont {Varvitsiotis}},\ and\ \bibinfo {author} {\bibfnamefont {Z.}~\bibnamefont {Wei}},\ }\bibfield  {title} {\bibinfo {title} {Completely positive semidefinite rank},\ }\href {https://doi.org/10.1007/s10107-017-1198-4} {\bibfield  {journal} {\bibinfo  {journal} {Mathematical Programming}\ }\textbf {\bibinfo {volume} {171}},\ \bibinfo {pages} {397} (\bibinfo {year} {2018})}\BibitemShut {NoStop}%
\bibitem [{\citenamefont {Maxfield}\ and\ \citenamefont {Minc}(1962)}]{maxfield-minc-1962}%
  \BibitemOpen
  \bibfield  {author} {\bibinfo {author} {\bibfnamefont {J.~E.}\ \bibnamefont {Maxfield}}\ and\ \bibinfo {author} {\bibfnamefont {H.}~\bibnamefont {Minc}},\ }\bibfield  {title} {\bibinfo {title} {On the matrix equation {$X'X=A$}},\ }\href {https://doi.org/10.1017/S0013091500014681} {\bibfield  {journal} {\bibinfo  {journal} {Proceedings of the Edinburgh Mathematical Society}\ }\textbf {\bibinfo {volume} {13}},\ \bibinfo {pages} {125} (\bibinfo {year} {1962})}\BibitemShut {NoStop}%
\bibitem [{\citenamefont {Motzkin}\ and\ \citenamefont {Straus}(1965)}]{motzkin-straus-1965}%
  \BibitemOpen
  \bibfield  {author} {\bibinfo {author} {\bibfnamefont {T.~S.}\ \bibnamefont {Motzkin}}\ and\ \bibinfo {author} {\bibfnamefont {E.~G.}\ \bibnamefont {Straus}},\ }\bibfield  {title} {\bibinfo {title} {Maxima for graphs and a new proof of a theorem of {Tur\'an}},\ }\href {https://doi.org/10.4153/CJM-1965-053-6} {\bibfield  {journal} {\bibinfo  {journal} {Canadian Journal of Mathematics}\ }\textbf {\bibinfo {volume} {17}},\ \bibinfo {pages} {533} (\bibinfo {year} {1965})}\BibitemShut {NoStop}%
\bibitem [{\citenamefont {Burer}\ \emph {et~al.}(2009)\citenamefont {Burer}, \citenamefont {Anstreicher},\ and\ \citenamefont {D{\"u}r}}]{burer-2009-dnn-cp}%
  \BibitemOpen
  \bibfield  {author} {\bibinfo {author} {\bibfnamefont {S.}~\bibnamefont {Burer}}, \bibinfo {author} {\bibfnamefont {K.~M.}\ \bibnamefont {Anstreicher}},\ and\ \bibinfo {author} {\bibfnamefont {M.}~\bibnamefont {D{\"u}r}},\ }\bibfield  {title} {\bibinfo {title} {The difference between {$5 \times 5$} doubly nonnegative and completely positive matrices},\ }\href {https://doi.org/10.1016/j.laa.2009.05.021} {\bibfield  {journal} {\bibinfo  {journal} {Linear Algebra and its Applications}\ }\textbf {\bibinfo {volume} {431}},\ \bibinfo {pages} {1539} (\bibinfo {year} {2009})}\BibitemShut {NoStop}%
\end{thebibliography}%

\end{document}